\documentclass[pre,twocolumn,showpacs,superscriptaddress]{revtex4-1}

\usepackage{graphicx}
\usepackage{color}
\usepackage{amsmath,amssymb}
\usepackage{mathrsfs}
\usepackage{caption}
\usepackage{subcaption}
\usepackage{threeparttable}
\usepackage{ragged2e}
\begin{document}

\title{ History effects on network growth}

\author{Hadiseh Safdari}
\affiliation{Department of Physics, Shahid Beheshti University, G.C., Evin,
Tehran 19839, Iran}
\author{Milad Zare Kamali}
\affiliation{Department of Physics, Shahid Beheshti University, G.C., Evin,
Tehran 19839, Iran}
\author{Amir Hossein Shirazi}
\affiliation{Department of Physics, Shahid Beheshti University, G.C., Evin,
Tehran 19839, Iran}
\author{ Moein Khaliqi}
\affiliation{Department of Mathematics, Tarbiat Modares University, P.O. Box 14115-175, Tehran, Iran}
\author{Gholamreza Jafari}
\affiliation{Department of Physics, Shahid Beheshti University, G.C., Evin,
Tehran 19839, Iran}

 \date{\today}

\begin{abstract}
Growth dynamic of real networks because of emerging complexities
is an open and interesting question.
Indeed it is not realistic to ignore history impact on the current events.
The mystery behind that complexity could be in the role of history in some how.
To regard this point, the average effect of history has been included by a kernel function
in differential equation of Barab$\acute{\textbf{a}}$si– Albert (BA) model .
This approach leads to a fractional order BA differential equation as
a generalization of BA model. As opposed to unlimited growth for
degree of nodes, our results show that over time the memory impact
will cause a decay for degrees. This gives a higher chance to
younger members for turning to a hub.
In fact in a real network, there are two competitive processes.
On one hand, based on preferential attachment mechanism nodes with higher
degree are more likely to absorb links. On the other hand, node history
through aging process prevents new connections.
Our findings from simulating
a network grown by considering these effects also from studying a
real network of collaboration between Hollywood movie actors conforms the results
and significant effects of history and time on dynamic.
\end{abstract}

\pacs{}

\maketitle

\section{Introduction}

One of the widely studied model to describe network dynamics,
Barab$\acute{a}$si- Albert (BA), is based on preferential attachment
mechanism \cite{Albert2,Dorogovtsev2,Albert3,Papadopoulos}. According
to this model, over time new nodes join system and link to the earlier
nodes, with probability proportional to the degree of
preexisting nodes. Nodes with higher degree have greater chance to
attract more new connections. This leads to the emergence of hubs with
very large degree. Therefore, BA model characterizes one of the
important features in real networks, namely, scale invariance
\cite{Jeong,Victor,Holme,Ravasz,Newman}.

BA model has been very successful in describing many properties of
real networks \cite{Albert2}, however, its unlimited growth prediction
for degree of nodes is not always in line with reality
\cite{Maru,Lehmann}. In fact, in real networks other factors work
against a node's growth which BA model fails to account for
\cite{Albert1,Medo,Lind}. For instance, in realistic networks like
citation and scientific collaboration network \cite{Leicht,Local}, world-wide web network \cite{Kumar}
and network of movie actors collaboration \cite{Watts}, there are some
important phenomena such as aging, screening and censorship  to be included
\cite{Adamic,Newman2,redner,Chechkin,scala,censor}.

In network of movie actors for example, overtime, hubs are replaced by
new ones.  Superstars get promoted for some time period as they get to
their height of popularity and eventually loose their attractiveness
in eyes of media and decay gradually. New generation of stars will
replace the old generation and repeat the same cycle again. An other
example is political network \cite{Burton,Chessa}, influence of people
in politics grow and eventually decreases over time, political hubs
are replaced by other people. No one experience unbound growth in
politics.

As opposed to many realistic situations where aging process causes the
hubs to be replaced by newcomers, in BA model an emergent hub will
always remain powerful and prevent newcomers from becoming strong hub.
In other word, we need to consider the effect of losing power for old
hubs and emergence of new hubs. Number of studies have been done in
this area and some strategies have been proposed
\cite{Albert1,Dorogovtsev,Chen,Sarshar}. As a common issue with most of
these approaches, the outcome has been requested manually from the model,
i.e, it is imposed on the network by adding some new terms to BA equation.
The objection against these approaches is that, new terms that they added are not capable to
demonstrate the main issue; even more, it is not always possible to
find a closed form solution.

According to BA model nodes can connect to each other without any
restriction, hence their degree can grow boundlessly. However, in
many real systems, each node has limited capacity for joining other
nodes. For example number of active friends a person can have is
limited as a person would denote part of his/her time to each of those
friends. Since we have limited amount of time per day/weak, one can
not simply have infinite friends.

A scientist or actor has just enough time to collaborate with small
number of partners. Therefore, having some connections makes the space
and opportunity scarce for more connections. It's current connections
do not allow the node to join other nodes easily. This is similar to
the phenomenon in fermi systems known as screening effect, reduction in
effective electronic potential because of the cloud of electrons
around one electron \cite{flensberg}. To extend the preferential
attachment mechanism, the model has to also pay attention to these
kind of effects.

\begin{figure*}
 \captionsetup{justification=raggedright,
singlelinecheck=false
}
 \begin{subfigure}{0.3\textwidth}
   \includegraphics[width=0.8\linewidth]{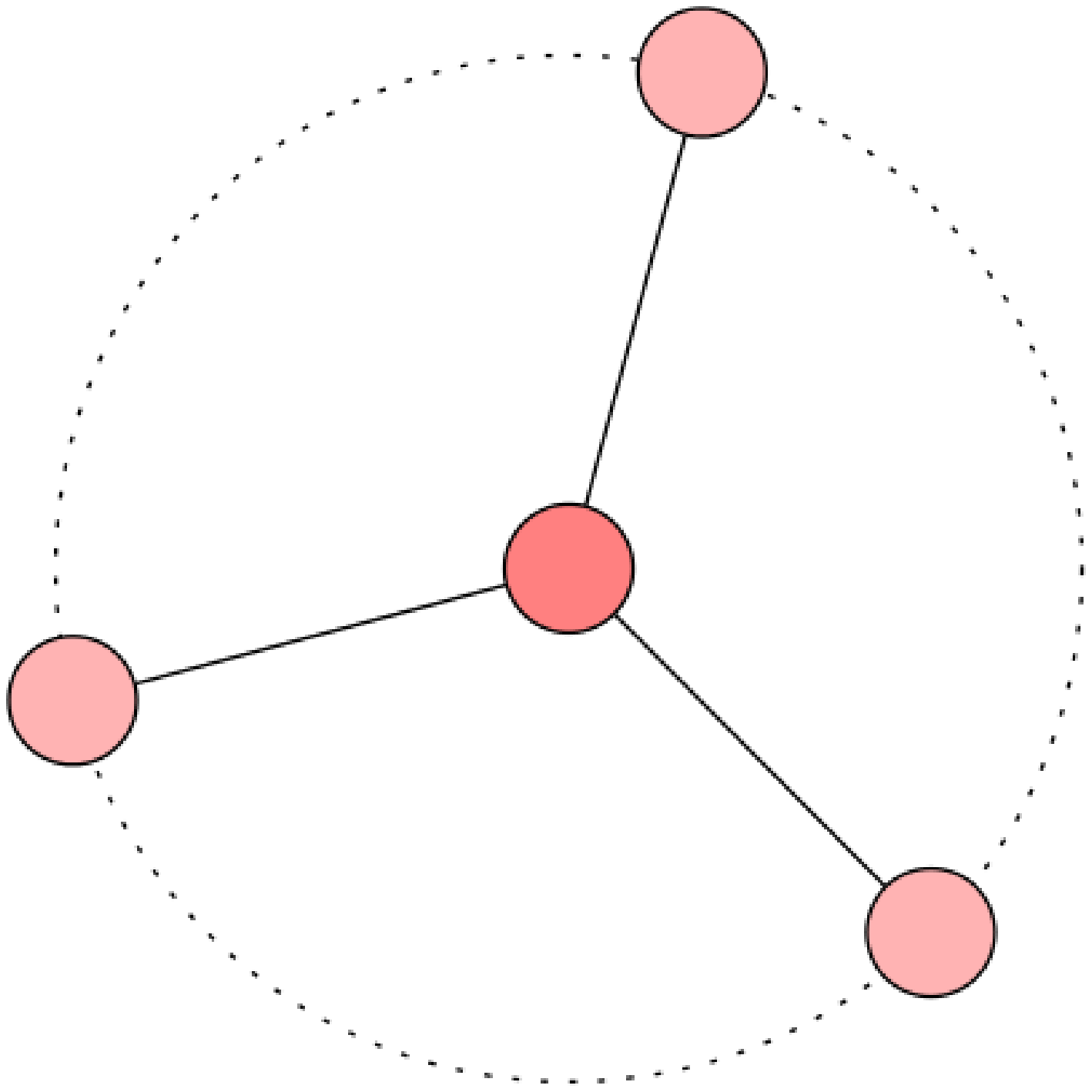}
   \caption{ }
   \label{plot_age0}
 \end{subfigure}
 \begin{subfigure}{0.33\textwidth}
   \hspace*{-0.0in}
   \includegraphics[width=0.8\linewidth]{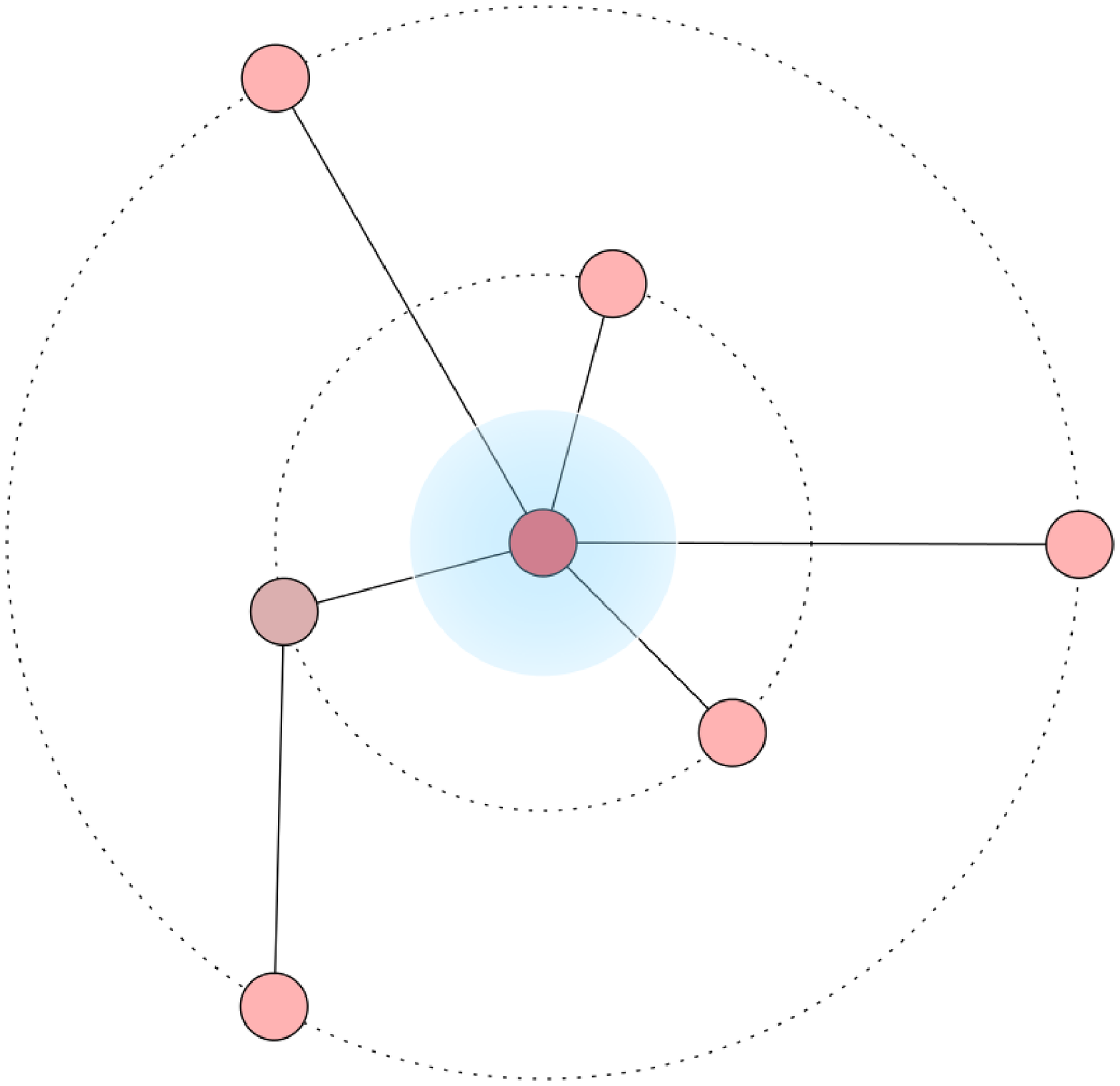}
   \caption{ }
   \label{plot_age1}
 \end{subfigure}%
 \begin{subfigure}{0.33\textwidth}
   \includegraphics[width=0.8\linewidth]{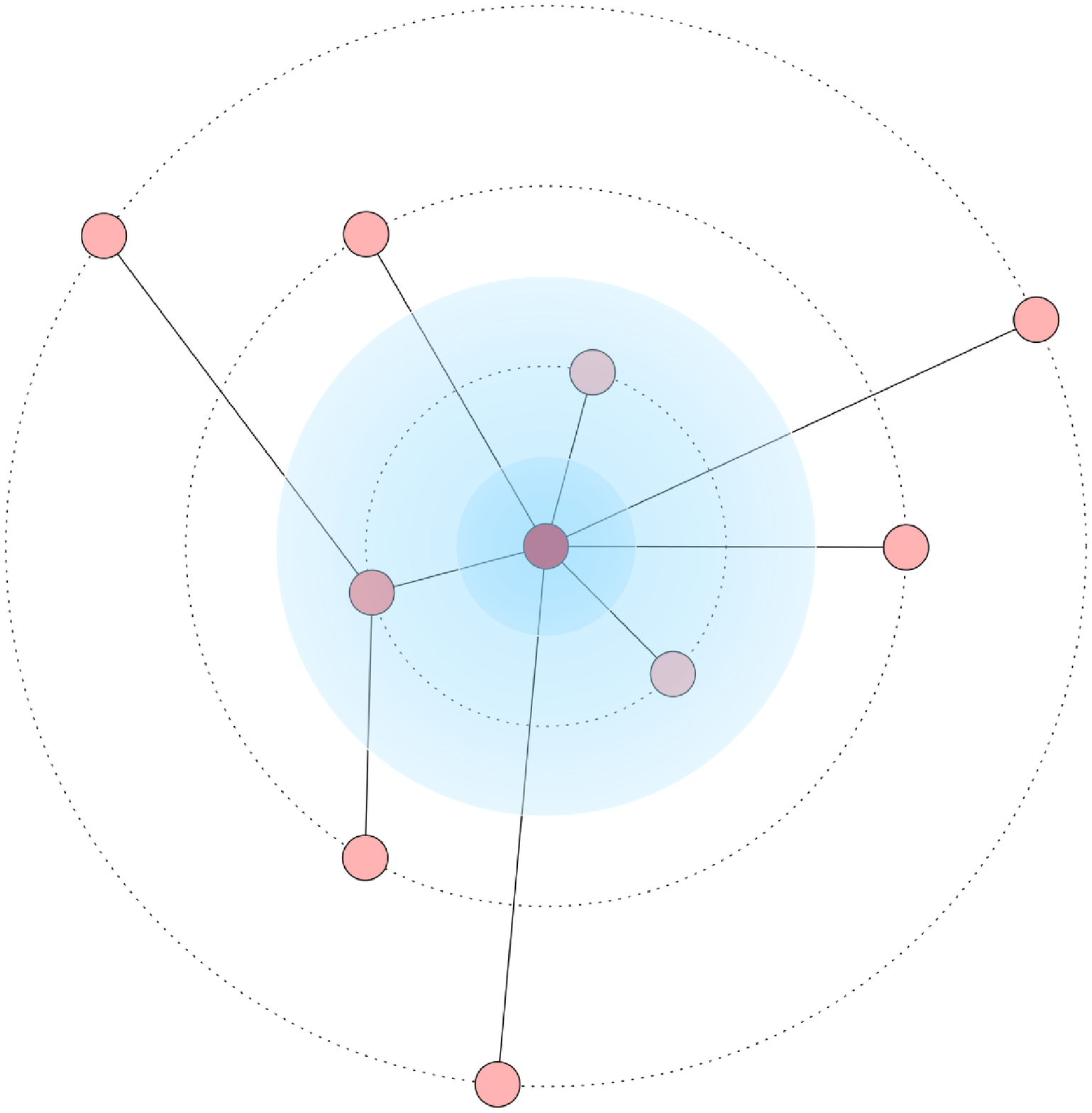}
   \caption{ }
   \label{plot_age2}
 \end{subfigure}
 \begin{subfigure}{0.3\textwidth}
  \includegraphics[width=0.8\linewidth]{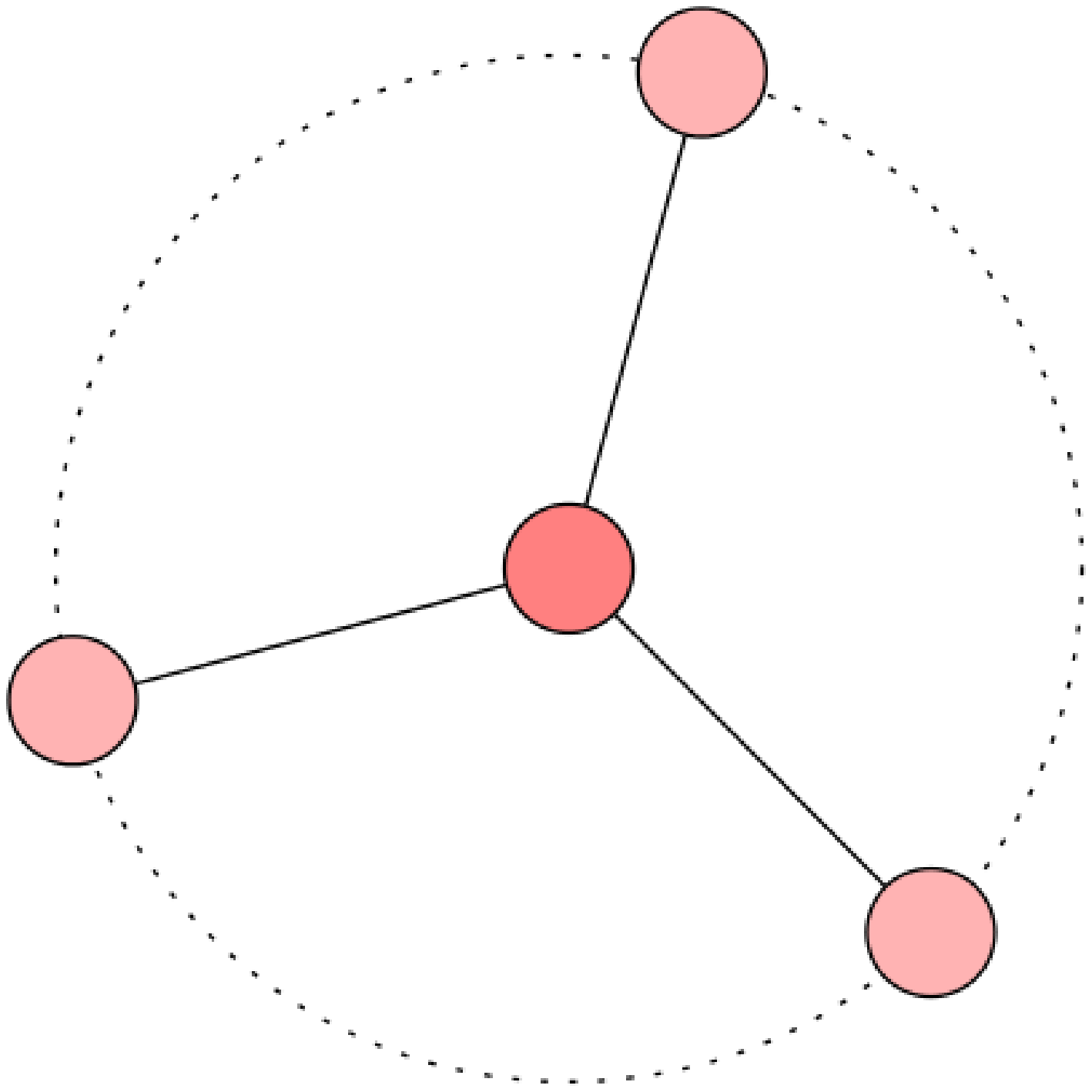}
   \caption{ }
  \label{plot_ba0}
 \end{subfigure}%
 \begin{subfigure}{0.33\textwidth}
   \includegraphics[width=0.8\linewidth]{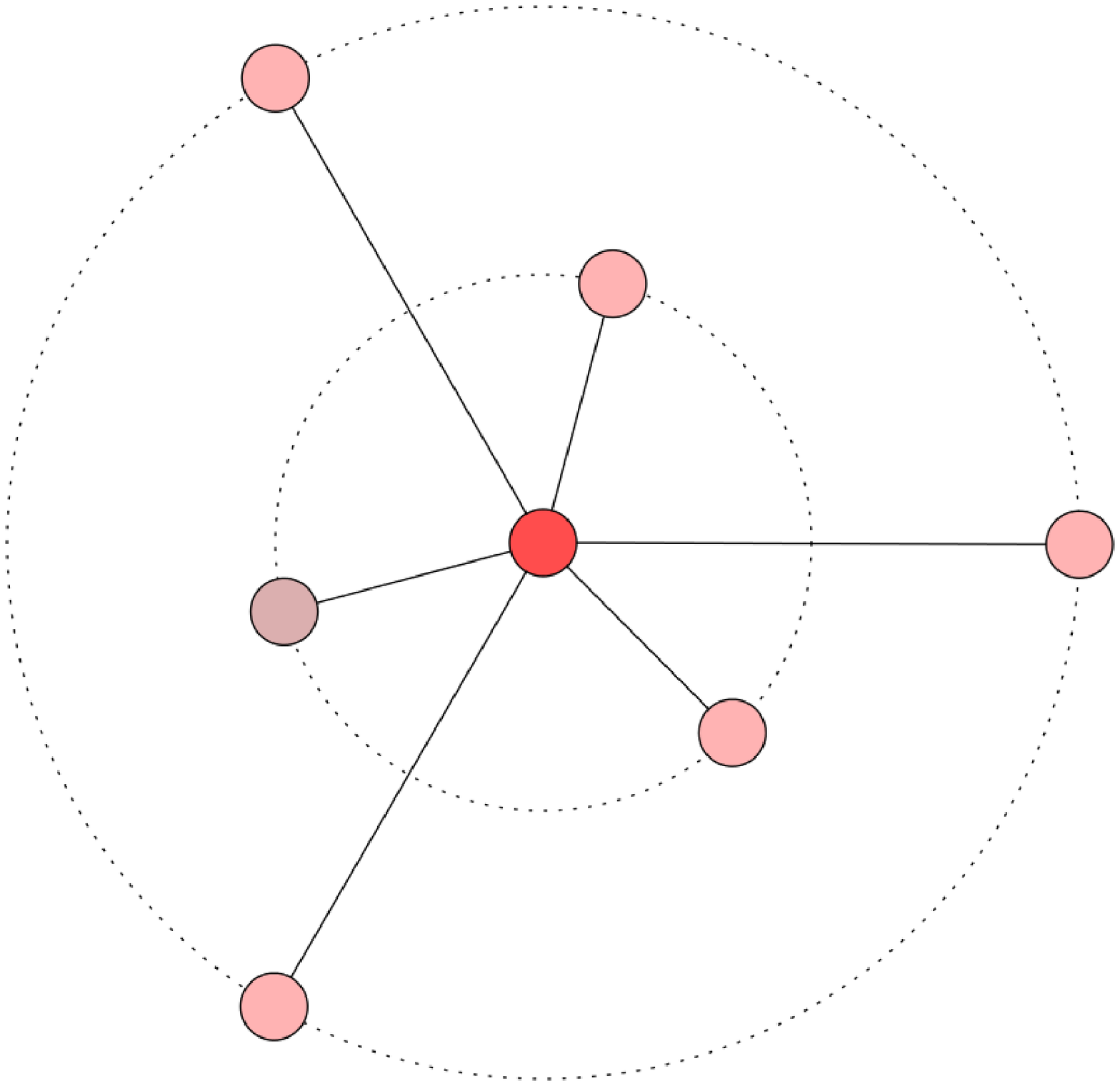}
   \caption{  }
   \label{plot_ba1}
 \end{subfigure}%
 \begin{subfigure}{0.33\textwidth}
   \includegraphics[width=0.8\linewidth]{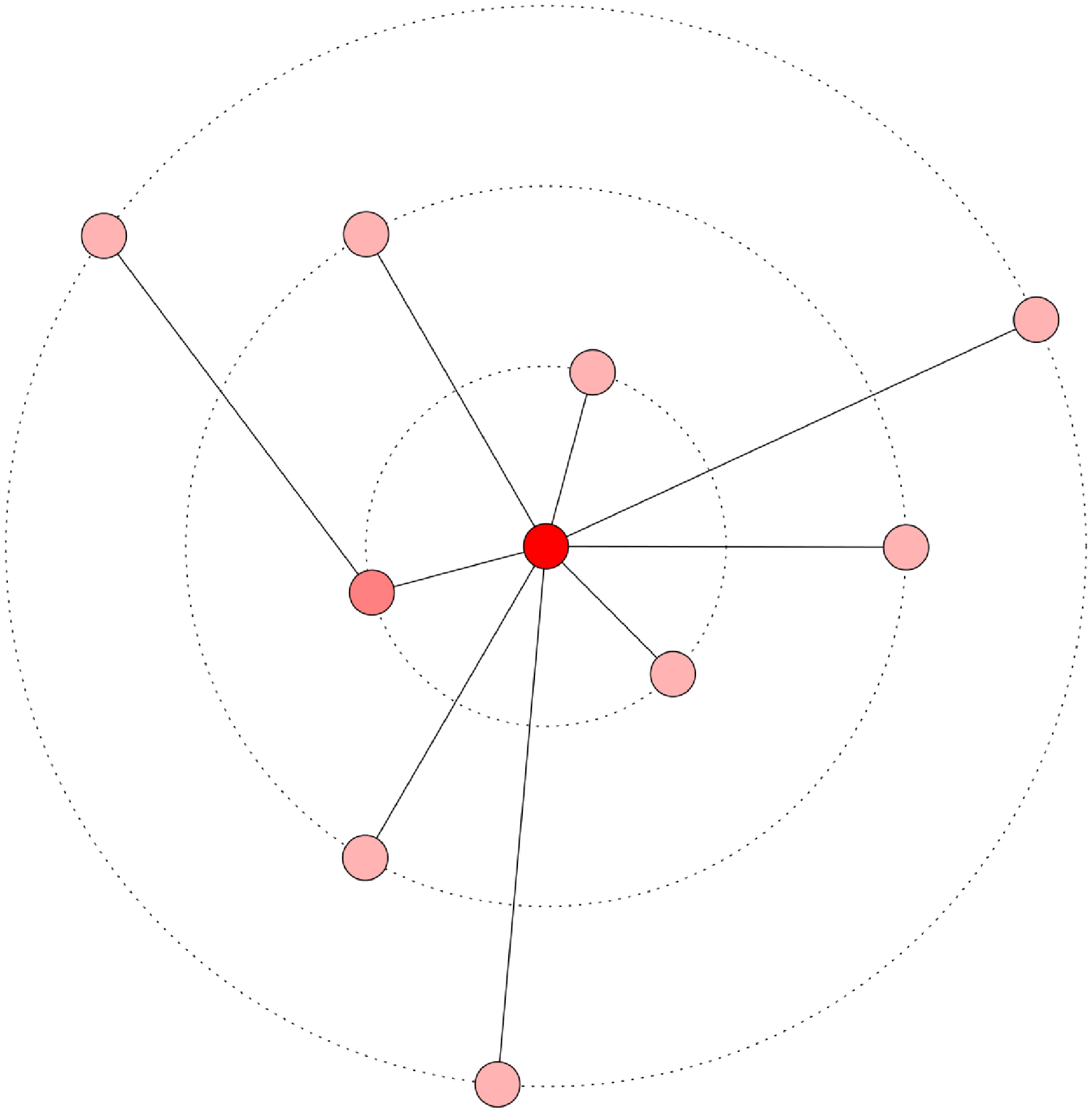}
   \caption{}
   \label{plot_ba2}
 \end{subfigure}
 \caption{ Schematic diagram represents dynamical growth of nodes in a network,
   developed under preferential attachment mechanism by considering limitations
   imposed by aging process and screening effect on growth process in
   (\ref{plot_age0}- \ref{plot_age2}) and
   comparison with standard preferential attachment without
   considering these restrictions (\ref{plot_ba0}- \ref{plot_ba2}).
   Red to blue color shades of nodes represent their attractiveness
   for new links.  Highest attractiveness is represented by red. The
   radial distance from central node shows arrival time for a
   node. Over the time a new node joins the network and connects to $m$
   existed nodes according to their degree. However, old members
   because of their age have lower chance of being selected. As
   (\ref{plot_ba0}- \ref{plot_ba2}) show, in the absence of these
   effects hubs are dominated in absorbing new links.}
 \label{fig:schem}
\end{figure*}
A good candidate to include history effects could be fractional
calculus. Fractional calculus, the generalized form of ordinary
differentiation and integration to non-integer order
\cite{Oldham,SAMKO,MILLER,Gorenflo,Klafter} has unique features such
as nonlocality and memory, making it highly applicable in many fields
of science and engineering \cite{BJWest,Grigolini,Philip,Lundstrom}. With
respect to the presence of kernel for history in fractional operators,
the results from this model carries the system memory. In other word,
memory of the system plays a substantial role in determining the
future path. The proposed approach in this work to generalize BA model
is based on fractional calculus. In this approach, the equation
governing the dynamic of network growth is a non integer order
differential equation. Hence, operators from fractional calculus
appear.

By working in fractional realm, the time distances are no longer
identical. In fact in the growth process, some nodes can face long
time delays before jumping to the next time step. In other words, they
become freezen in time. At the same time some other nodes experience
very short waiting period and pass more time steps relative to the
frozen ones.

A simulated network based on fractional order BA differential equation
approach has also been developed in which an old node looses
effectiveness of its degree. Therefore its attractiveness to absorb new links will
decrease over time. Figure (\ref{fig:schem}) schematically represents
dynamical growth of a network evolved based on preferential attachment
mechanism. Moreover, aging and screening effects have been
considered. As a node gets older, its probability to attain more links
from new members decreases. Existing hubs will decay eventually and
new hubs can emerge.

The rest of this paper is organized as the following. In section
(\ref{section1}) fractional version of BA model by use of fractional
calculus operators is introduced. Section (\ref{section2}) devotes to
solving the fractional equation and it's results. Results of a network
simulation by considering aging effects on it's dynamic has been
presented and discussed in section (\ref{section3}). In section
(\ref{section4}) the real network of Hollywood movie actor collaboration
which shows bounded growth for collaboration rate is investigated.
Section (\ref{sec:Conclusion}) concludes the results.

\section{Fractional approach for preferential attachment algorithm}
\label{section1}

In the BA model for network growth, the governing equation system,
\begin{eqnarray}
\label{barabashi}
\begin{array}{lr}
\frac{dk_{i}(t)}{dt}= \frac{mk_{i}(t)}{\sum_{i}^{t} k_{j}(t)}   \\
k_{i}(t_{i0}) = m,
\end{array}
\end{eqnarray}
is an integer order equation for each node $i$, does not include the past history of the
considered node. Only node's degree in previous step affects its
present degree. In order to consider the history on the growth
process, a kernel on time, $\kappa (t-t')$, will be imposed on the
right hand side of the Eq. (\ref{barabashi}). This kernel accounts for
the memory \cite{West,MAINARDI,Goychuk,Metzler}.

\begin{eqnarray}
\label{kernel1}
\frac{dk_{i}(t)}{dt}=\int_{t_{0}}^{t}  dt'  \kappa (t-t')  \frac{mk_{i}(t')}{\sum_{j}^{t} k_{j}(t)}.
\end{eqnarray}

In case of memoryless systems, kernel should be a Dirac delta
function, $\delta(t-t')$ which results in standard differential
equation in BA model for network growth.
Here, we choose a power law functional form for kernel on time,
capable of involving the past. This kernel,
guarantees existence of scaling feature as it is an intrinsic nature
of most phenomena \cite{Hansen,Mantegna,peng}. A distant past event should have far less effect on
present, compared to near past events, only exceptions are large
influential events which even over shadow recent ordinary
events. Substituting power law kernel in (\ref{kernel1}), it becomes:
\begin{eqnarray}
\label{kernel2}
\frac{dk_{i}(t)}{dt}=\frac{1}{\Gamma(\alpha-1)}\int_{t_{0}}^{t} dt'
(t-t')^{\alpha-2} \frac{mk_{i}(t')}{\sum_{j}^{t} k_{j}(t)}.
\end{eqnarray}
From fractional calculus it is apparent that the right hand side of
this equation is a fractional integral of order $(\alpha-1)$,
$_{t_{0}}D_{t}^{-(\alpha-1)}$ on the interval $[t_{0},t]$
\cite{MILLER}. Therefore, it can be shown as,
\begin{eqnarray}
\label{kernel3}
\frac{dk_{i}(t)}{dt}=_{t_{0}}D_{t}^{-(\alpha-1)} \left[ \frac{mk_{i}(t)}{\sum_{j}^{t} k_{j}(t)} \right].
\end{eqnarray}

Now applying a fractional Caputo derivative of order $(\alpha-1)$
\cite{caputo} on both side of the above equation, we can write it in
the form of a differential equation,
\begin{eqnarray}
\label{kernel4}
 _{t_{0}}^{c}D_{t}^{\alpha}\left[ k_{i}(t)\right]  = \frac{mk_{i}(t)}{\sum_{j}^{t} k_{j}(t)}.
\end{eqnarray}

For a continuous function $f$ on the $[a,b]$ interval, left Caputo
derivative of order $\alpha$ is defined as follow \cite{torres}:
\begin{eqnarray}
\label{eqf-1}
\begin{array}{lr}
^{c}_{a}D_{t}^{\alpha}\left[f(t) \right]= \\
\qquad \qquad \quad\frac{1}{\Gamma(n-\alpha)}\int_{a}^{t}(t-\xi)^{n-\alpha-1}(\frac{d}{d\xi})^{n}f(\xi)d\xi, \\
\end{array}
\end{eqnarray}
where $n$ is the smallest integer greater than or equal to $\alpha$,
$n=\left[ \alpha \right]+1$. Caputo derivative has the advantage that
in solving fractional differential equations (FDE), uses integer order
boundary or initial conditions.

In the last step to obtain Eq. (\ref{kernel4}), we applied the fact
that Caputo fractional derivative and fractional integral are inverse
operators \cite{torres2}, for $\alpha > 0$,
\begin{eqnarray}
\label{kernel5}
 _{t_{0}}^{c}D_{t}^{\alpha}\left[ _{t_{0}} D_{t}^{-\alpha} \right] f(t) = f(t).
\end{eqnarray}

In this way, the governing equation is a fractional order
differential equation guarantees the presence of memory.

\section{numerical results of fractional order growth equation}
\label{section2}

\begin{figure*}[t]
\captionsetup{justification=raggedright,
singlelinecheck=false
}
   \centering
  \begin{subfigure}{0.5\textwidth}
    \centering
    \includegraphics[width=0.9\linewidth]{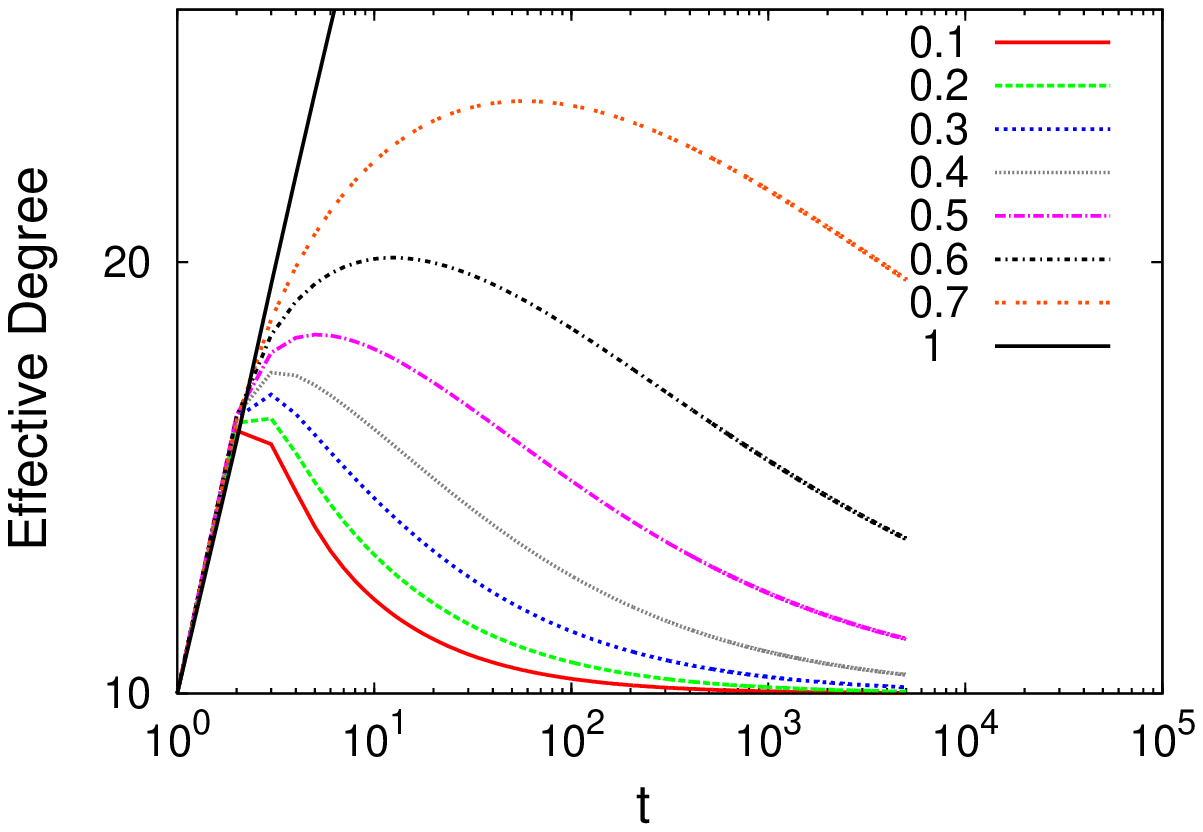}
    \caption{}
    \label{fig:num1}
  \end{subfigure}%
  \begin{subfigure}{0.5\textwidth}
    \centering
    \includegraphics[width=0.9\linewidth]{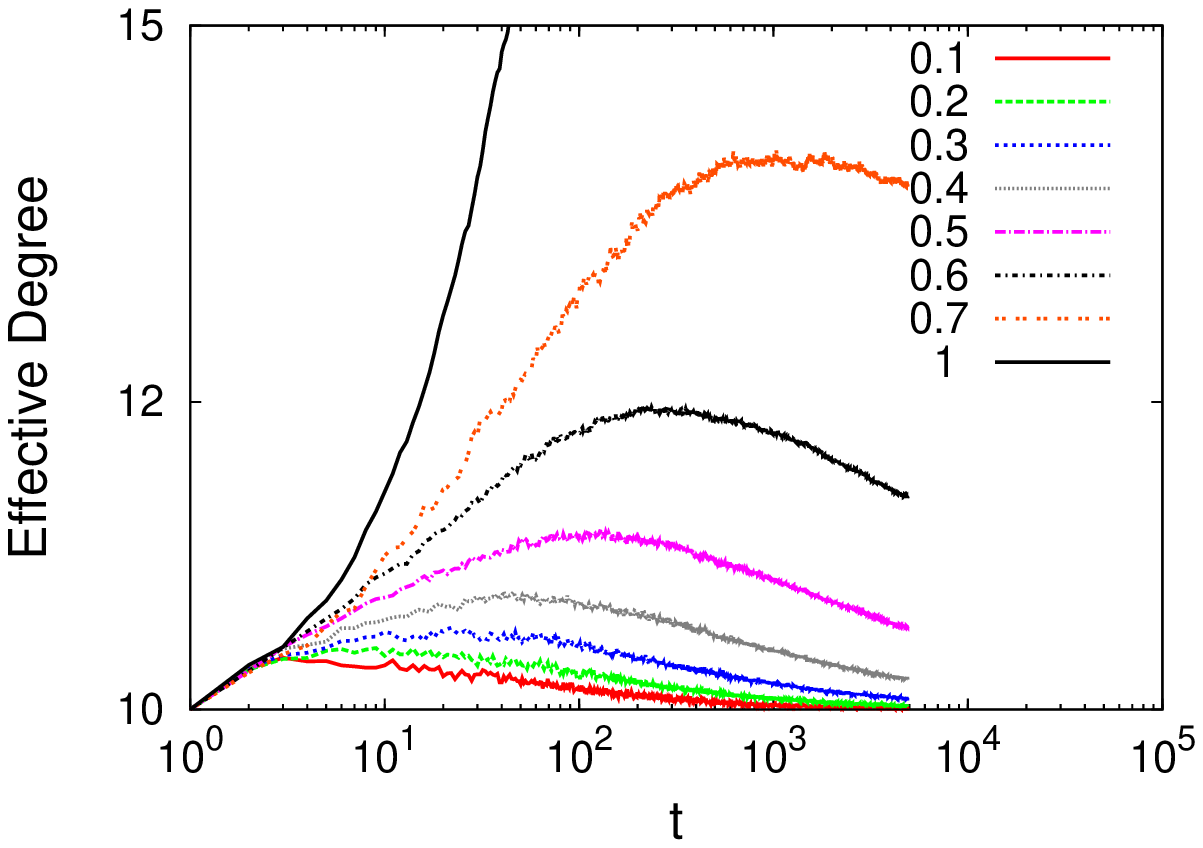}
    \caption{}
    \label{fig:sim2}
  \end{subfigure}%
\rightmark
\caption{ Figure (\ref{fig:num1}) is numerical solution of Eq.
  (\ref{eq_num}) and (\ref{fig:sim2}) is mean degree of nodes from
  simulating a network that it's growth dynamic is affected by aging
  process. It is clear from results that k(t) behaves differently from
  unlimited growth predicted by BA model. Due to aging process k(t)
  reaches a peak and declines gradually afterwards. Simulation has
  been done for $5000$ time step with initial $m_{0}=10$ nodes and
  every new node connects to $m=10$ earlier nodes.}
\label{simulation}
\end{figure*}
In attempting to deploy fractional calculus in BA model of growing
networks, we start by dynamic equation in form of,
\begin{eqnarray}
\label{frac_ba}
\begin{array}{lr}
^{c} _{t_{0}} D_{t}^{\alpha} k_{i}(t) = m \frac{k_{i}(t)}{\sum_{j}^{t} k_{j}(t)},   \\
k_{i}(t_{i0}) = m.
\end{array}
\end{eqnarray}
Where $0< \alpha \leq 1$. For $\alpha=1$ the above equation becomes
the well-known dynamic equation in BA model. We have a system of fractional order
differential equations coupled by the summation in denominator.
From here the problem becomes an initial value problem for FDE,
\begin{eqnarray}
\begin{array}{lr}
\label{fde}
^{c} _{t_{0}} D_{t}^{\alpha} y(t) = f(t,y(t)),   \\
y(t0) = y_{0},
\end{array}
\end{eqnarray}
which can be solved numerically by the predictor-corrector algorithm
\cite{Lubich,Diethelm,Garrappa}. Hence, we can reformulated it to
equivalent Volterra integral equation in the form,
\begin{eqnarray}
\label{Voltera}
y(t)=y_{0}+\frac{1}{\Gamma({\alpha})}\int_{t_{0}}^{t}(t-s)^{(\alpha-1)}f(s,y(s))ds.
\end{eqnarray}

To deal with this integral, we use product rectangle method, which
divides the domain into $n$ fragments, $t_{j}=t_{0}+jh$, with equal
space $h$. The right hand side function, in terms of numerical
approximation $y_{j}$ for $y(t_{j})$, is denoted by
$f(t_{j},y_{j})$. Finally, we have the discrete form as follows,
\begin{eqnarray}
\label{disy}
y_{n}=y_{0}+h^{\alpha} \Sigma_{j=0}^{n-1} b_{n-j-1}f_{j},
\end{eqnarray}

with $b_{n}$ coefficients as,
\begin{eqnarray}
\label{bcoeff}
b_{n}= \frac{(n+1)^{\alpha}-(n)^{\alpha}}{\Gamma(\alpha+1)}.
\end{eqnarray}

As a result, the discrete form of fractional order differential
equation (\ref{frac_ba}) becomes,
\begin{eqnarray}
\label{disk}
k_{n}=k_{0}+h^{\alpha} \Sigma_{j=0}^{n-1} b_{n-j-1} \frac{m k_{j}}{\sum_{j}^{t} k_{j}(t)}.
\label{eq_num}
\end{eqnarray}

Here, $b_{n}$'s are time dependent coefficients which indicate the
aging effect. This factor shows contribution of the previous degree of
$i$th node on its present value. With increasing the lifetime (increasing
$n$), $b_{n}$ becomes smaller. In other words, over the time old links of a node
will loose their effect on its growth process based on $b_{n}$'s coefficients.
Therefore, the effective degree of a node (Eq. (\ref{eq_num})) will decrease. For $\alpha=1$ the $b_{n}$ converges
to unity and we get back to the standard BA model in which old degrees
all have the same weight.
 By numerically solving equation system (\ref{disk}), we found interesting
 results confirming the effect of history on network evolution. The
 results in Fig. (\ref{fig:num1}) for various amounts of $\alpha$,
 show time dependence of $k_{i}(t)$ (effective degree has been left for $i$th node at time $t$). It can be
 seen that for all values of $\alpha < 1$, effective degree of node $i$
 increases at first. It reaches the maximum value, then starts to
 decay.

 This behavior could be as a result of competition between two mechanisms governing the
 network dynamic. On one hand, nodes absorb new links according to
 preferential attachment algorithm, on the other hands, aging process
 will reduce the probability of being selected by new members by reducing its effective degree at that moment. The
 older the member, the smaller is its portion from past history. However, on the
 same time period, node with higher degree would receive less impact from
 aging than node with smaller degree.

Many real world networks exhibit the above mentioned behavior. Here
as a well known example which will be studied in particular in section
(\ref{section4}), network of movie actors collaboration could be
mentioned. In this network, we can see that superstars do collaborate
with many actors as they become popular. However, because of screening
effect one can just have the ability to cooperate with finite number
of actors. There is not enough room for collaboration with every
one. The rate of collaboration will reach a peak and then decline
gradually over time as a result of aging process.

\begin{figure}[b]
\captionsetup{justification=raggedright,singlelinecheck=false}
\includegraphics[width=7.cm,height=5.5cm,angle=0]{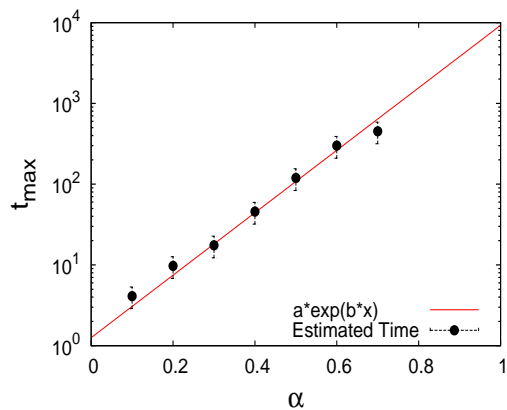}
\caption{The time to reach maximum for different $0 < \alpha < 1$
  values with the fitted curve. This time decreases exponentially with
  increasing the exponent.  By considering even a very small effect of
  memory on evolution, which means any deviation from $\alpha=1$,
  there would be a peak after that degree of nodes will experience a
  decay. However, for closer values to $1$ this time tends to
  infinity.}
\label{peak}
\end{figure}

The growth and decaying rates depend on the order of fractional
equation.  Smaller $\alpha$ reaches the peak quickly but larger
$\alpha$ takes more time.  The smaller the exponent is, the more
aging can reduce growth rate. Getting closer to exponent $1$ aging
effect is reduced and nodes have longer growth time. For $\alpha=1$
which results in preferential attachment without aging, it increases
boundlessly. Exponential increase in time distance that effective degree takes to reach
its maximum for different $\alpha$ has been shown in Fig. (\ref{peak}).

\begin{figure*}[t]
\captionsetup{justification=raggedright,singlelinecheck=false}
  \centering
    \begin{subfigure}{0.33\textwidth}
    \centering
    \includegraphics[width=6.cm,height=4.5cm]{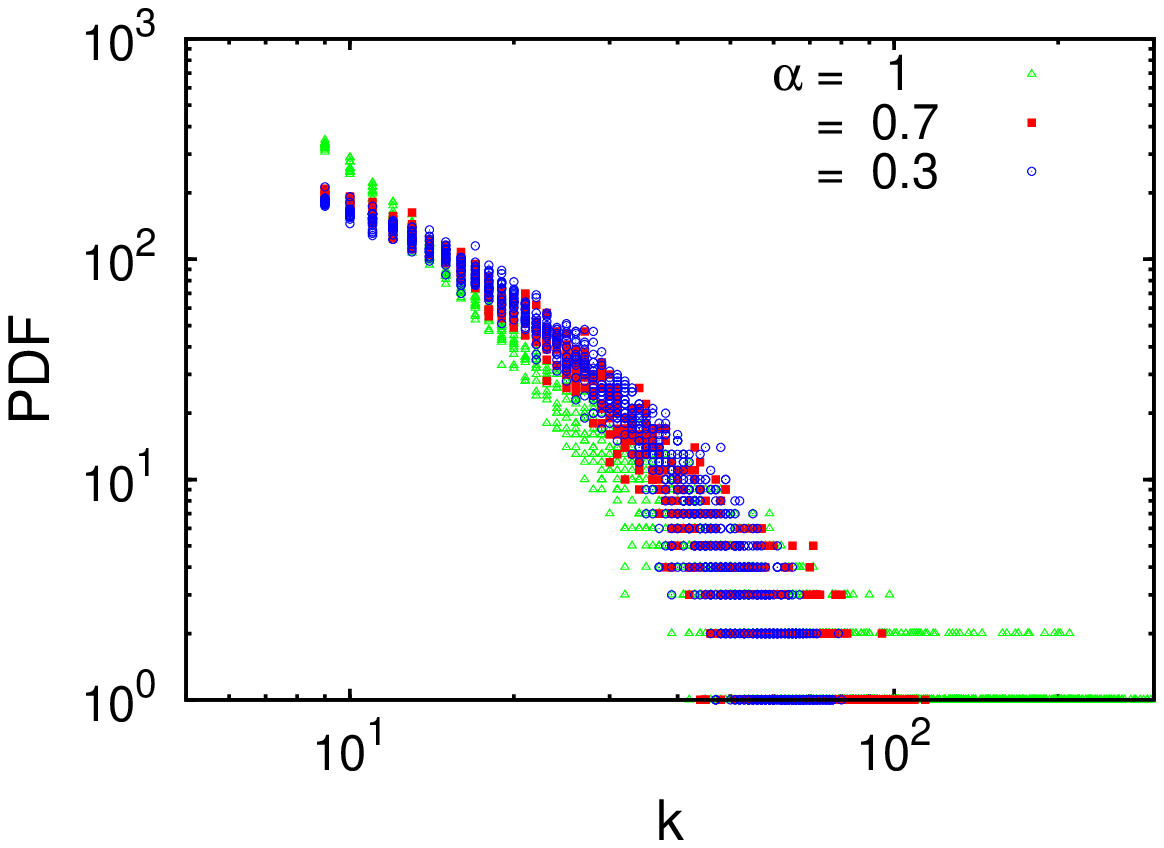}
    \caption{ }
    \label{plot_hista}
  \end{subfigure}
  \begin{subfigure}{0.33\textwidth}
    \centering
    \includegraphics[width=6.cm,height=4.5cm]{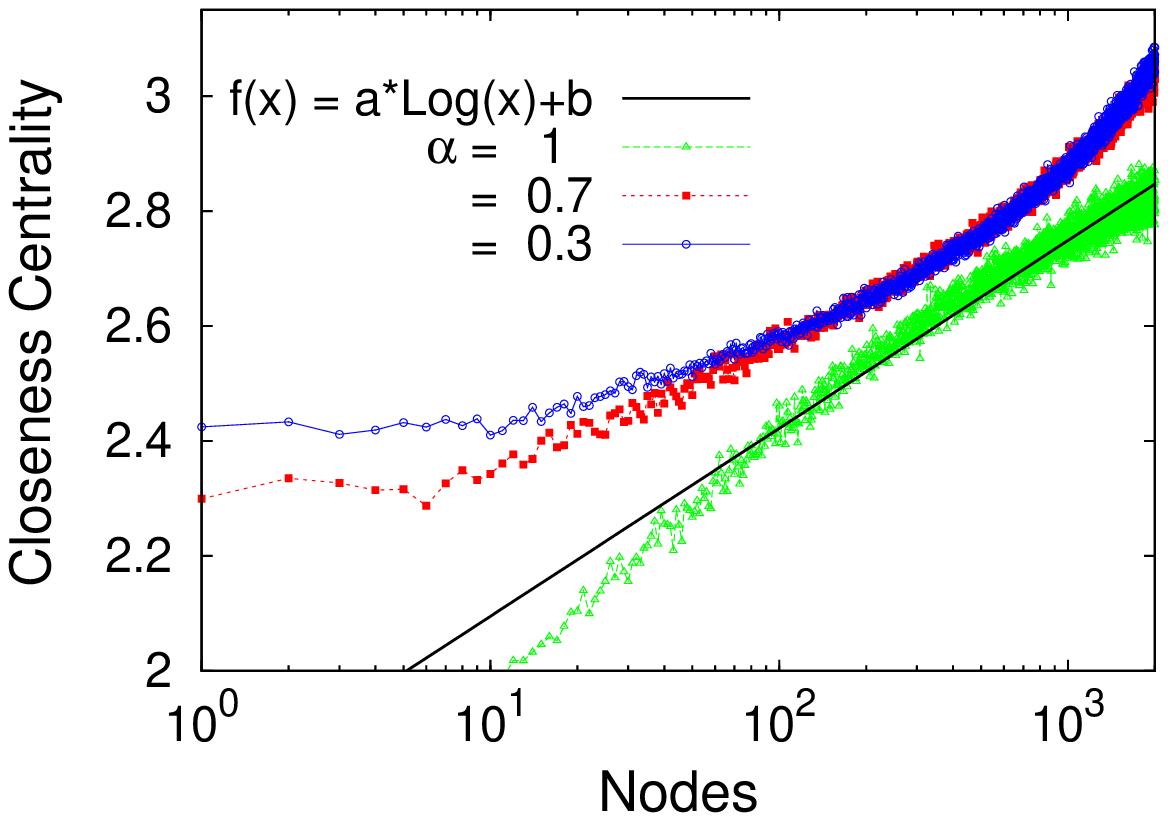}
    \caption{ }
    \label{plot_meana}
  \end{subfigure}%
  \begin{subfigure}{0.33\textwidth}
    \centering
    \includegraphics[width=6.cm,height=4.5cmh]{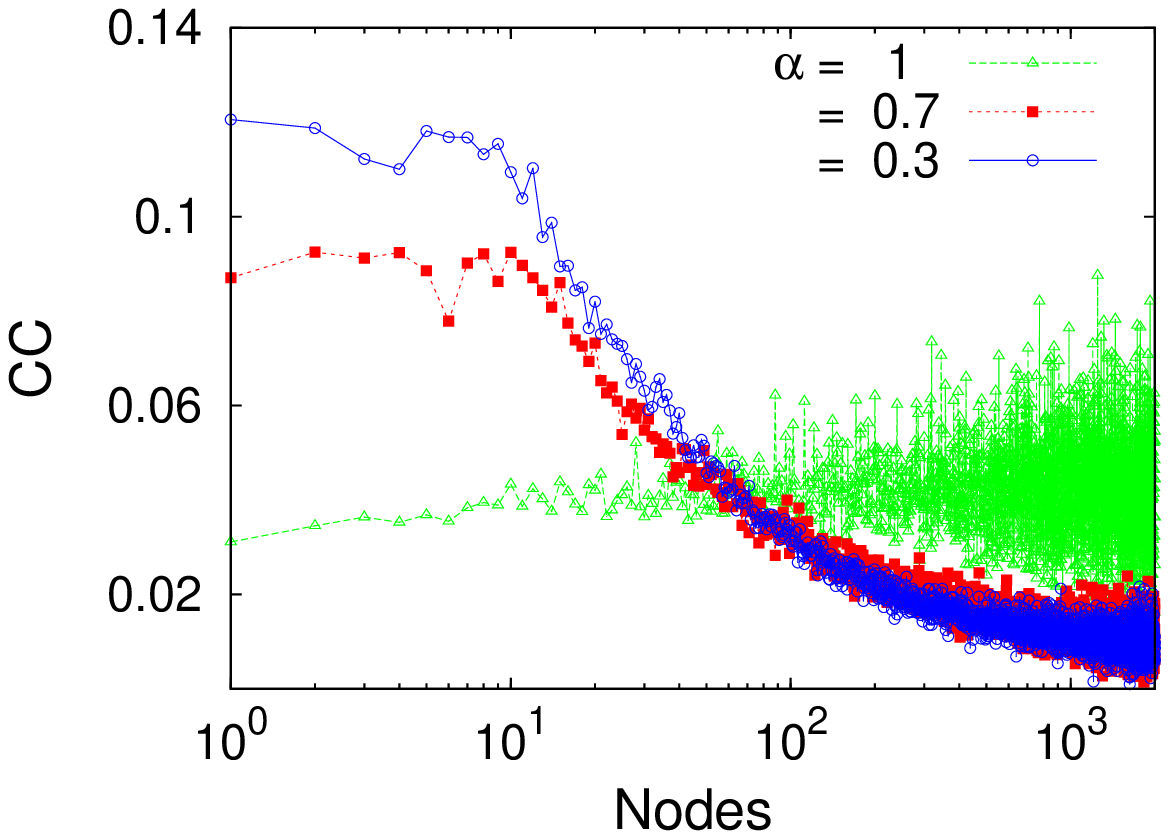}
    \caption{ }
    \label{plot_clust}
  \end{subfigure}
  \begin{subfigure}{0.35\textwidth}
    \centering
    \includegraphics[width=6.75cm,height=4.5cm]{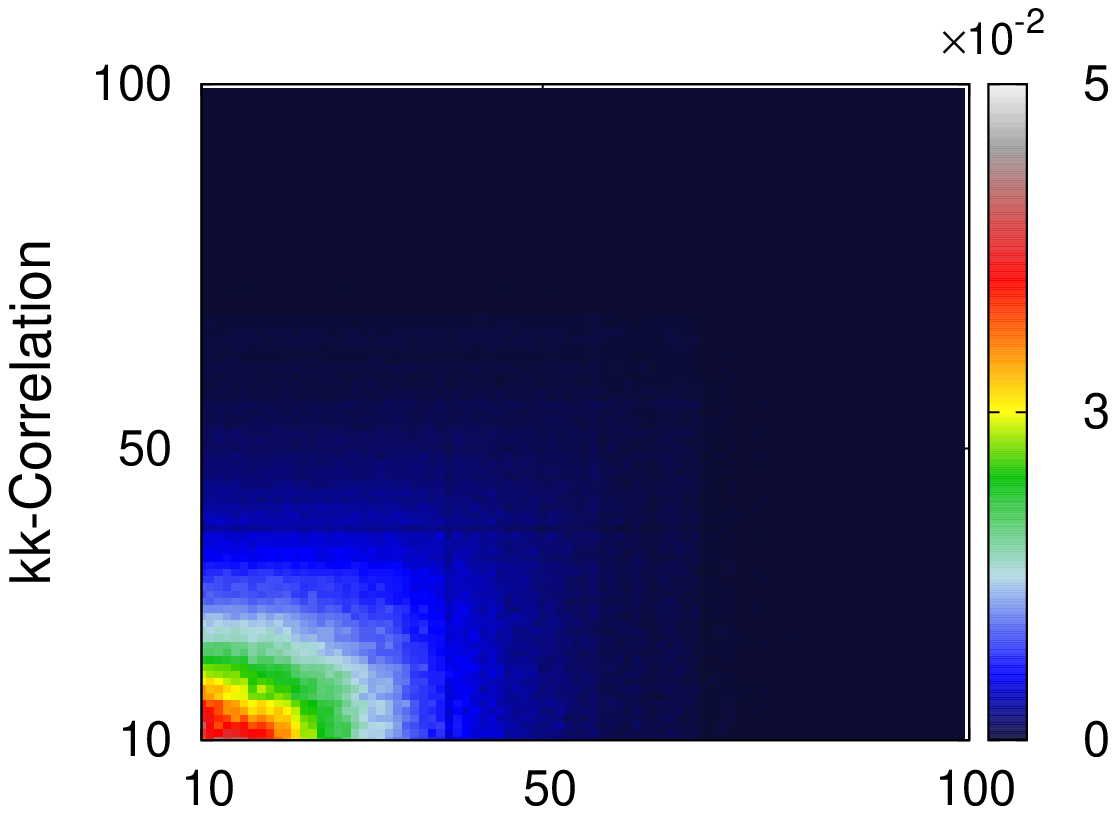}
    \caption{$\alpha =0.3$}
    \label{plot_core_3}
  \end{subfigure}%
  \begin{subfigure}{0.35\textwidth}
    \centering
    \includegraphics[width=6.cm,height=4.5cm]{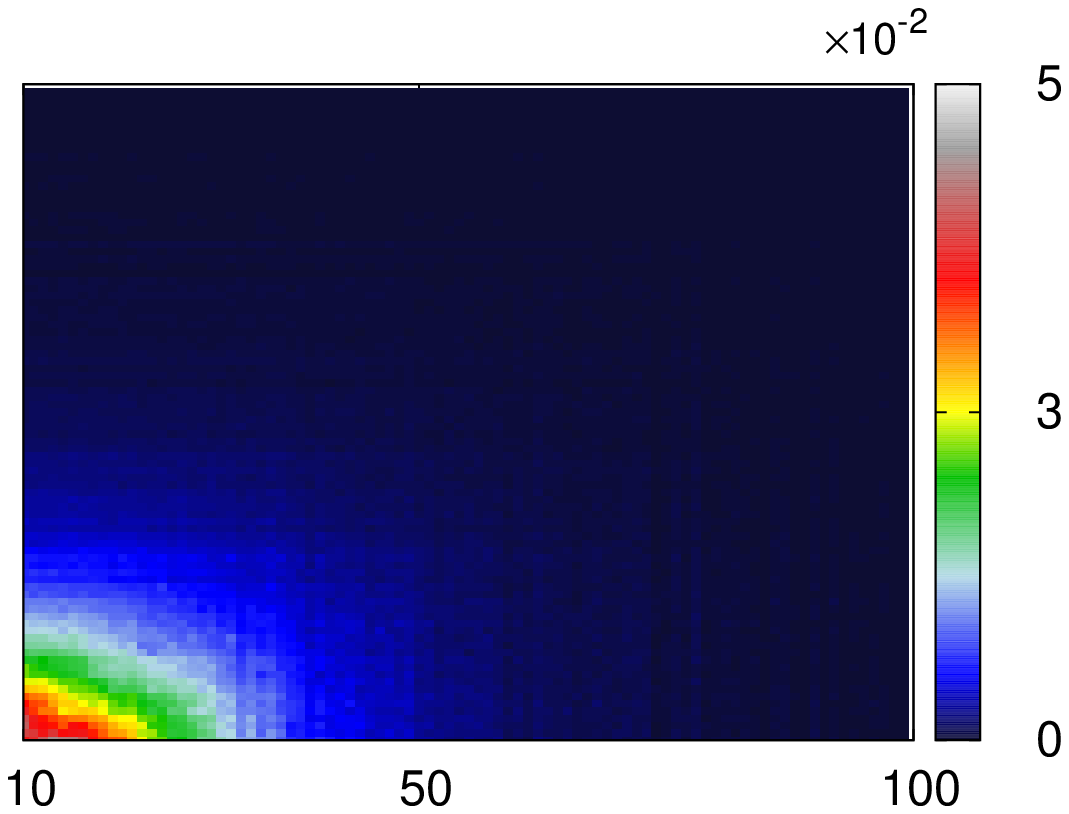}
    \caption{$\alpha =0.7$ }
    \label{plot_core_7}
  \end{subfigure}%
  \begin{subfigure}{0.35\textwidth}
    \centering
    \includegraphics[width=6.cm,height=4.5cm]{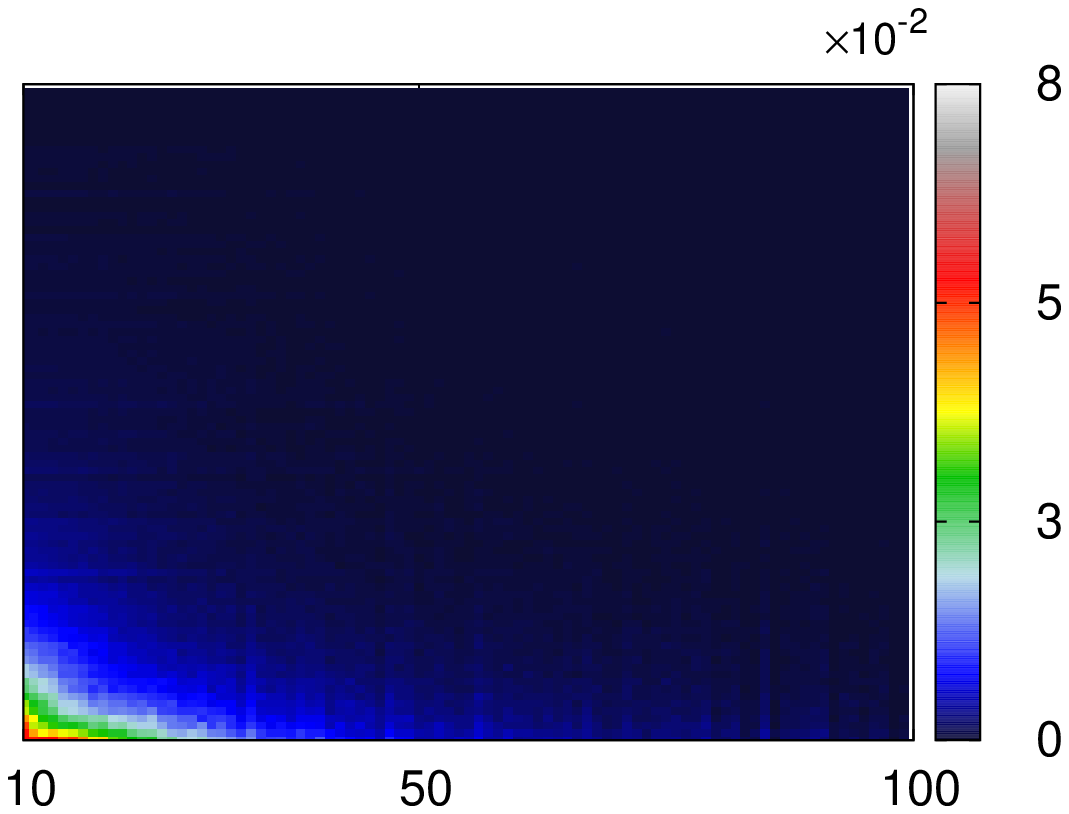}
    \caption{$\alpha = 1$ }
    \label{plot_core_10}
  \end{subfigure}
\caption{(\ref{plot_hista}) shows probability distribution function
  for different orders of fractional derivative. While for $\alpha=1$,
  BA model, the expected power law behaviour is found, for fractional
  orders less than unity, there is deviation from power law. This
  deviation gets severe for smaller $\alpha$. It could be caused by
  two competitive processes, namely preferential attachment mechanism
  and screening effect. (\ref{plot_meana}) is closeness centrality
  averaged over all nodes; comparison between three exponents
  $\alpha=.3,0.7,1$. Horizontal axis is nodes according to their
  arrival time.  For higher values of $\alpha$, this measure has
  higher amounts. Elders have shortest path to others.
  (\ref{plot_clust}) indicates clustering coefficient averaged over
  all nodes for three exponents $\alpha=.3,0.7,1$. Horizontal axis
  shows nodes according to their arrival time. As it is clear, it has
  certain behavior. Older nodes have more communities around.
  (\ref{plot_core_3}-\ref{plot_core_10}) show degree-degree
  correlation averaged over all nodes for three exponents
  $\alpha=.3,0.7,1$. Tendency to bind with similar members have been
  increased by including history effects. The fact is that the
  inclination to powerfuls is reduced. Powerful sites would not be in
  power for all the time. There is a competition by new
  generations. Hence, it is not the priority anymore to connect to the
  elders.}
  \label{fig:image2}
\end{figure*}
Consequently, the fractional order BA equation for growing networks
can demonstrate the existence of history. Because fractional order
derivatives by applying a kernel over time involve the effect of
elapsed time, it indicates the fact that every member has an end and
will be isolated. In many real networks, members do not remain in the
system for ever. As time passes, they will be set aside from the
community. The results from fractional equation (\ref{disk}) remarks
this fact.

\section{Simulating an aged Network}
\label{section3}

Along side solving fractional order dynamic equation governing the network, we have simulated a network
with aging process. In this simulation we start by a fully connected
network with $m+1$ members, hence each node has initial degree of $m$.
In each time step $i$, node $i$ is introduced to the network, which is
linked to $m$ previous nodes proportional to their degree.
\begin{figure*}[t]
  \captionsetup{justification=raggedright, singlelinecheck=false }
\centering
      \includegraphics[width=12cm,height=7.5cm,angle=0]{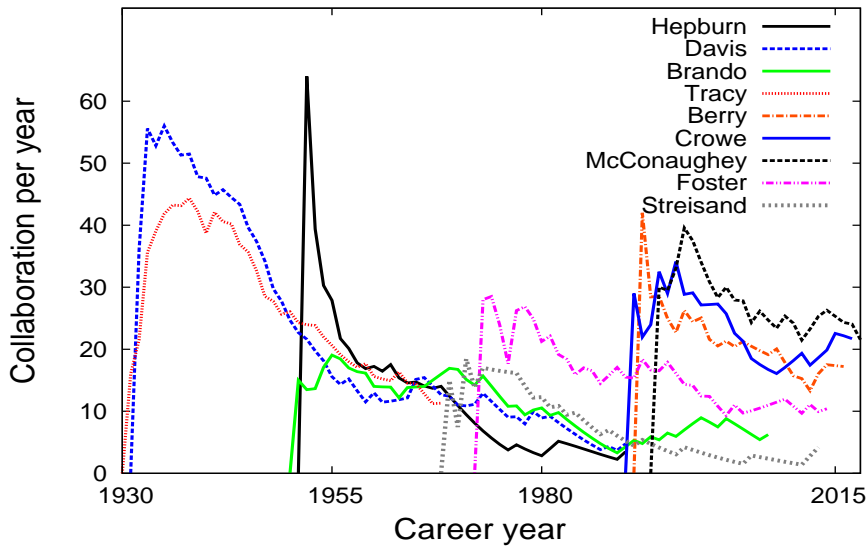}
      \caption{A sample of total collaboration per year for Oscar
        winner actors.  Collaboration rate is number of staring
        actors/actress a actor plays with in a movie each year. It is
        clear that each actor would experience a peak and gradual
        decay in their total collaboration as they age
        professionally.}
\label{plot_actors}
\end{figure*}
To include aging process in the growth dynamic, the effectiveness of
old links for a node is lessened. A link between two nodes, is less
effective for old node, however it preserves its influence for younger
one; similar to a connection between two nodes in a directed weighted
network.  As a consequence, the probability of being selected by newly
arrived members will be reduced for members with more old links. This
reduced effectiveness of older links in growth dynamic has been
applied by using $b_n$ coefficients which are proportional to link's
life time as weight factors in calculating probability of receiving
new links.

Figure (\ref{fig:sim2}) displays time dependence of effective degree
on time steps resulted from simulation. The observed behavior is in
good agrement with numerical results, (Fig. (\ref{fig:num1})),
obtained from fractional order differential equation
(\ref{disk}). Just like numerical solution, initially, effective
degree per time of a node increases, after reaching its maximum value
it decays gradually.

In figure (\ref{fig:image2}), we have shown some statistical
proprieties of the network simulation with aging concept along side BA
model. Small $\alpha$ represents severe aging effects, therefore
behavior of quantities with smaller $\alpha$ shows greater deviation
from BA model.

Degree distribution reveals major characteristics of a model. For
preferential attachment it follows a power law form \cite{Albert2},
however in presence of aging it shows deviation from power law. This
deviation from power law could be as a result of competition between
aging process and preferential attachment mechanism. According to
preferential attachment nodes with high degree absorb most of new
links, however, getting older over the time reduces probability of
high degree nodes to be selected \cite{Lehmann}, since aging reduces
the effective degree of nodes.  This reduces growth rate of older
nodes which gives nodes with smaller degree a higher chance to receive
new links.

The average shortest path to all nodes in a network \cite{Freeman}
known as closeness centrality is a criteria to measure power of centrality for nodes in network.
This quantity has been studied for different orders of
fractional BA equation. It can be seen from Fig. (\ref{plot_meana})
that closeness centrality for nodes has raised for aged system
compared to BA model. This figure shows that older nodes in BA model
have more central position in network respect to olders in aged system. Moreover, in an aged system,
older members are more close to others than younger nodes.

Aging process also has impacts on clustering coefficient of nodes,
Fig. (\ref{plot_clust}). It is an indicator that displays tendency of
a node's neighbors to connect to each other \cite{Watts}. Senior
members are associated with communities which have high degree of
connectivity as opposed to recently added members. Although aged nodes
lose their importance, they still preserve their strategic position
among others. However, in BA model older nodes have smaller clustering
coefficient than olders in system by including aging effect.

One significant consequence of considering aging is that new members
can have the opportunity to grow and become a hub, as opposed to BA
model where only early members have a chance of fast growth. Hence
tendency of nodes to connect to their similar nodes, what is called
assortativity \cite{Newman4} will increase. This can be seen in
degree-degree correlation palette in
Figs. (\ref{plot_core_3}-\ref{plot_core_10}). Despite BA model that
considerable connections are with hubs or older nodes, in aged
networks remarkable correlation can be found between similar nodes in
degree.

\section{Network of Oscar winners collaboration}
\label{section4}

\begin{figure}[t]
\captionsetup{justification=raggedright,
singlelinecheck=false}
\centering
      \includegraphics[width=8cm,height=5.5cm,angle=0]{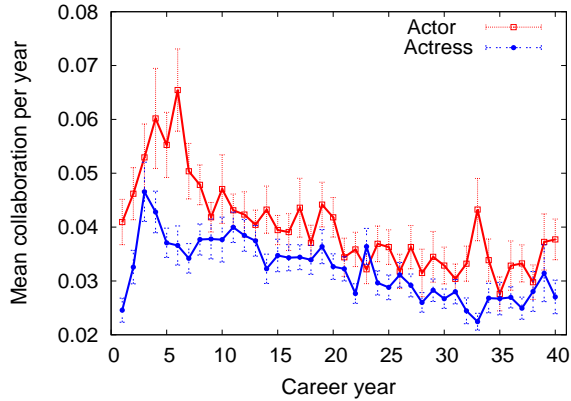}
\caption{Averaged number of collaboration per year for $75$ Oscar winner actors
   and $74$ actresses. The rate of collaboration in both curves increases
  and after reaching a maximum value, it will decay gradually.}
\label{plot_ave_actors}
\end{figure}

In order to verify the above results for a real network, collaboration
network of Hollywood movie actors have been studied. We extracted list
of actors from the Internet Movie Database available at www.imdb.com.
The table of all movies and star actors of those movies have been
collected. TV series and TV movies are not included in our data. Nodes
in the intended network are actors and they have a common link if they
have acted in the same movie.  As a case study in this network, total number of
collaboration of Oscar winners with other actors per year has been
derived as shown in figure (\ref{plot_actors}) for some of them. It
is the total number of staring actors that an actor/actress played with
in each year. It can be seen that almost all the actors presented in
the above network have the same pattern in their career. At start
of acting career, collaboration rate is increased till it reaches a
maximum. After that, collaboration rate will decrease
gradually. Despite some exceptions, it is a general behavior that
collaboration rate will decrease ultimately. The average
collaboration rate of $75$ actors and $74$ actresses show the mentioned pattern in the network,
Fig. (\ref{plot_ave_actors}).  Although the general behavior is
similar for both curves, the mean value for number of
collaboration for actresses reaches the maximum sooner. It behaves as
if its dynamic equation has smaller order of differentiation,
$\alpha$, than the curve for actors collaboration. In other words,
history effects and aging process have more severe impacts on
actresses growth. Besides, studying the mean rate of collaboration on
first and second half of the whole period shows no noticeable difference
in dynamical behavior.  The above results are in remarkable agreement
with findings of the numerical and simulations in the above sections
as shown in Figure (\ref{simulation}).

\section{Conclusion}
\label{sec:Conclusion}

Complexity and collective behavior are key characteristics of many
natural and social systems. The past of a realistic systems can not be
ignored in its dynamics and what is happening now. Even it could be
said, since future expectations of a system are reselections of system
goals, a complex system dynamic can be characterized by its goals too.
Some mathematical models have been proposed to expand BA model by
adding new terms to it's differential equation, however, since these
additional terms are not unique, there is no clear method to prefer
one over the other. More over often these additional terms do not
often have analytical solution.

In attempt to describe how the history play role in the current events
and growth dynamic of networks, we apply a kernel function as a
average of the history on the standard equation in BA model. This
choice for the kernel shows interwoven of past events that leads to
dynamic of the system. This approach leads to a fractional order BA
differential equation governing network dynamic.

According to results, this approach predicts a boundary for growth of
members. Just by generalizing governing equation in BA model to an
equation with fractional order derivative, a more realistic dynamic
for systems is achieved.  As it can be seen in many real systems such
as movie actors collaboration, networks of scientific papers,
friendship networks and etc, degree of nodes progress for a certain
period of time and then decay gradually after that. This change in
topology of the system causes redistribution in the members power.

Moreover, even future goals of a system be it minimizing system energy
or adaptability with nature (in evolution) could shape its dynamic
too. Future projection of kernel could be tough of as the way that
future goal might effect one's present decisions. Birds for example
migrate seasonally to survive their life in next seasons. Therefore,
considering the future goals could be an interesting study in network
dynamics.


\end{document}